\documentclass[aps,pre,preprint,superscriptaddress,nofootinbib]{revtex4-1}
\usepackage{amsmath}
\usepackage{fullpage}
\usepackage{amssymb}
\usepackage{amsthm}
\usepackage{hyperref}
\usepackage{setspace}
\usepackage{graphicx}
\usepackage{bm}
\usepackage{physics}
\usepackage{float}

\begin{document}

% Use the \preprint command to place your local institutional report
% number in the upper righthand corner of the title page in preprint mode.
% Multiple \preprint commands are allowed.
% Use the 'preprintnumbers' class option to override journal defaults
% to display numbers if necessary
%\preprint{}

%Title of paper
\title{Stochastic path integrals can be derived like quantum mechanical path integrals}

% repeat the \author .. \affiliation  etc. as needed
% \email, \thanks, \homepage, \altaffiliation all apply to the current
% author. Explanatory text should go in the []'s, actual e-mail
% address or url should go in the {}'s for \email and \homepage.
% Please use the appropriate macro foreach each type of information

% \affiliation command applies to all authors since the last
% \affiliation command. The \affiliation command should follow the
% other information
% \affiliation can be followed by \email, \homepage, \thanks as well.
\author{John J. Vastola}
\email[john.j.vastola@vanderbilt.edu]{}
\affiliation{Department of Physics and Astronomy, Vanderbilt University, Nashville, Tennessee}
\affiliation{Quantitative Systems Biology Center, Vanderbilt University, Nashville, Tennessee}

\author{William R. Holmes}

%\homepage[]{Your web page}
%\thanks{}
%\altaffiliation{}
\affiliation{Department of Physics and Astronomy, Vanderbilt University, Nashville, Tennessee}
\affiliation{Quantitative Systems Biology Center, Vanderbilt University, Nashville, Tennessee}
\affiliation{Department of Mathematics, Vanderbilt University, Nashville, Tennessee}

%Collaboration name if desired (requires use of superscriptaddress
%option in \documentclass). \noaffiliation is required (may also be
%used with the \author command).
%\collaboration can be followed by \email, \homepage, \thanks as well.
%\collaboration{}
%\noaffiliation

\date{\today}

\begin{abstract}
Stochastic mechanics---the study of classical stochastic systems governed by things like master equations and Fokker-Planck equations---exhibits striking mathematical parallels to quantum mechanics. In this article, we make those parallels more transparent by presenting a quantum mechanics-like formalism for deriving a path integral description of systems described by stochastic differential equations. Our formalism expediently recovers the usual path integrals (the Martin-Siggia-Rose-Janssen-De Dominicis and Onsager-Machlup forms) and is flexible enough to account for different variable domains (e.g. real line versus compact interval), stochastic interpretations, arbitrary numbers of variables, explicit time-dependence, dimensionful control parameters, and more. We discuss the implications of our formalism for stochastic biology. 
\end{abstract}

% insert suggested keywords - APS authors don't need to do this
%\keywords{}

%\maketitle must follow title, authors, abstract, and keywords
\maketitle

\section{Introduction}
\label{sec:intro}

There is a compelling analogy between quantum mechanics and the burgeoning field of \textit{stochastic mechanics} \cite{baez2018}---the study of classical stochastic systems governed by things like master equations and Fokker-Planck equations. Although the path integral in quantum mechanics \cite{feynman1948} is sometimes interpreted as saying something profound about quantum particles exploring all possible paths at once \cite{dobson2000}, it turns out that path integrals are also useful tools in the decidedly less quantum realms of chemical kinetics \cite{elgart2004, elgart2006, li2016twoscale, benitez2016, thomas2014}, gene regulation \cite{wang2011, li2014, li2016landscape, li2013, wang2010, newby2014}, population dynamics \cite{spanio2017, kamenev2008, tauber2011, tauber2012, serrao2017}, and neuron firing \cite{ocker2017, iomin2019, baravalle2017, bressloff2015, bressloff2014hybrid, qiu2018, buice2013}, just to name a few. 

In this article we will concern ourselves with the path integral associated with the Fokker-Planck equation, a partial differential equation (PDE) that describes the time evolution of the probability density function for certain stochastic systems (like gene regulation models in the regime that molecule concentrations can be taken to be continuous variables \cite{gillespie2000}). It is interesting, and perhaps a little unfortunate, to note that this so-called `stochastic' path integral is usually derived in a completely different way than through the usual elementary quantum mechanics argument (to be briefly reviewed in Sec. \ref{sec:review}). This unnecessarily hides an intimate link between the Schr\"{o}dinger equation and its associated path integral on the one hand, and the Fokker-Planck equation and \textit{its} associated path integral on the other hand. Moreover, the Doi-Peliti \cite{doi1976, doi1976second, peliti1985} path integral description of master equation dynamics---which reduces to the Fokker-Planck path integral in the appropriate limit \cite{weber_master_2017}---is usually derived in the same way as the quantum mechanical path integral (again, see Sec. \ref{sec:review} for a brief review), so having a quantum mechanics-like formalism available would help clarify the relationship between the two path integrals. 

% Comment on imaginary time quantum mechanics here. Note that this is sort of like imaginary time quantum mechanics with state-dependent mass and momentum-dependent potentials. 

Existing derivations argue by (i) using the infinitesimal transition probability and exploiting the Markov property of these stochastic processes \cite{ZinnJustin2002, TongNotes, weber_master_2017, Graham1973}; (ii) writing the probability of a specific path through state space in terms of many delta functions and then using their usual integral representation \cite{chow_path_2015, Bressloff2014, altland_simons_2010}; or (iii) deriving the Fokker-Planck path integral from some other path integral \cite{weber_master_2017, kleinert2009}\footnote{See, for example, Ch. 18 (``Nonequilibrium Quantum Statistics'') of Kleinert \cite{kleinert2009}, where the Onsager-Machlup path integral associated with Brownian motion is derived from the path integral of a quantum mechanical particle coupled to a thermal reservoir.} depending on the context. The infinitesimal transition probability derivation involves cumbersome Jacobian calculations, while the delta function method is hard to generalize to PDEs other than the Fokker-Planck equation.

%Similarly, one can argue that the choice $\hat{p} := - i \partial/\partial x$ from quantum mechanics and Brownian motion is justified by...

Here, we will present a method for deriving the Fokker-Planck equation's path integral that closely mimics the derivation familiar from elementary quantum mechanics, and that does not suffer from the aforementioned problems. The associated formalism has the added benefit of being easy to understand, more flexible (for example, it is able to account for different domains, different stochastic interpretations, arbitrary numbers of variables, explicit time-dependence, and more) than the previously mentioned approaches, and mechanical---one is able to just `turn the crank' for a large variety of stochastic systems and construct a path integral. 

Our work also clarifies how the appropriate path integral description of a problem depends upon the domain of the underlying variables (e.g. real line versus half-line versus compact interval), and suggests a natural notion of a `momentum' operator. For example, the choice $\hat{p} := - \partial/\partial x$ (used in population dynamics \cite{kamenev2008, assaf2013, eriksson2013} and chemical kinetics \cite{escudero2009, hnatic2013}) can in some sense be linked to the idea that the natural `conjugate' problem for master equation and Fokker-Planck dynamics defined on $[0, \infty)$ is the Laplace transformed analogue.

This paper will proceed as follows. In Sec. \ref{sec:review}, we will remind the reader how the quantum mechanics and Doi-Peliti path integrals are derived, to make their connection to the present derivation more clear. In Sec. \ref{sec:vtypes}, we spell out our assumptions about the stochastic dynamics we are trying to describe. In Sec. \ref{sec:1Dderivation}, we present our derivation of the path integral in an illustrative case. In Sec. \ref{sec:modifications}, we present different ways our formalism could be altered or extended to describe different kinds of stochastic dynamics. Finally, we discuss some consequences for stochastic biology in Sec. \ref{sec:discussion}, and conclude in Sec. \ref{sec:conclusion}.

\section{Review of quantum mechanics and Doi-Peliti path integrals} \label{sec:review}

In this section, we will review the argument usually used to derive the path integral in elementary quantum mechanics, along with the argument Peliti used in his seminal 1985 paper \cite{peliti1985} to derive the Doi-Peliti path integral. The purpose of this section is to make the connection between our stochastic dynamics derivations and these two canonical derivations more transparent. 

\subsection{Quantum mechanics path integral derivation}

In this subsection, we will review the usual derivation \cite{schwartz2014, shankar2011, shankar2017} of the quantum mechanical path integral from the Schr\"{o}dinger equation. For general-purpose references on quantum mechanical path integrals, see Feynman and Hibbs \cite{feynman2010}, Schulman \cite{schulman1981}, and Kleinert \cite{kleinert2009}. 

%General references: CITE FEYNMAN AND HIBBS, KLEINERT

%Typical derivation references: \cite{schwartz2014}, KLEINERT?, Ch. 21 of SHANKAR PRINCIPLES (mentions both config and phase), Ch. 5 of SHANKAR QFT (mentions both config and phase)

For simplicity, consider a single particle in one dimension subject to a (time and momentum-independent) potential $V$ on the infinite line $(-\infty, \infty)$. The Schr\"{o}dinger equation reads
\begin{equation} \label{eq:schrodinger}
\begin{split}
i \hbar \frac{\partial \psi(x,t)}{\partial t} &= \hat{H} \psi(x, t) \\
&= \left[ \frac{\hat{p}^2}{2m} + V(\hat{x}) \right] \psi(x, t) \\
&= - \frac{\hbar^2}{2m} \frac{\partial^2 \psi(x,t)}{\partial x^2} + V(x) \psi(x, t) \ ,
\end{split}
\end{equation}
where $\hat{H}$ is the Hamiltonian operator, $\hat{p}$ is the momentum operator, and $\hat{x}$ is the position operator. 

In Dirac's bra-ket notation, the Schr\"{o}dinger equation reads
\begin{equation} \label{eq:schrodingerbraket}
i \hbar \frac{\partial}{\partial t} \ket{\psi} = \hat{H} \ket{\psi} \ ,
\end{equation}
which can be formally solved via 
\begin{equation} \label{eq:SEformalsln}
\ket{\psi(t_f)} = e^{-  \frac{i}{\hbar} \hat{H} (t_f - t_0)} \ket{\psi(t_0)}
\end{equation}
where $\ket{\psi(t_0)}$ is the system's state at initial time $t_0$, and $\ket{\psi(t_f)}$ is the system's state at a later time $t_f$. \\

To construct the path integral, two resolutions of the identity are required:
\begin{equation} \label{eq:QMxROI}
1 = \int_{-\infty}^{\infty} dx \ \ket{x} \bra{x} 
\end{equation}
\begin{equation} \label{eq:QMpROI}
1 = \frac{1}{2 \pi \hbar} \int_{-\infty}^{\infty} dp \ \ket{p} \bra{p} \ .
\end{equation}

The formal solution can be rewritten suggestively by inserting many position eigenket resolutions of the identity (Eq. \ref{eq:QMxROI}):
\begin{equation} \label{eq:SEsandwich}
\begin{split}
\ket{\psi(t_f)} =& e^{- \frac{i}{\hbar} \hat{H} \Delta t} e^{- \frac{i}{\hbar} \hat{H} \Delta t} \cdots e^{- \frac{i}{\hbar} \hat{H} \Delta t} \ket{\psi(t_0)} \\
=&\int dx_0 dx_1 \cdots dx_N \ket{x_N}  \\
& \times \matrixel{x_N}{e^{- \frac{i}{\hbar} \hat{H} \Delta t}}{x_{N-1}} \matrixel{x_{N-1}}{e^{- \frac{i}{\hbar} \hat{H} \Delta t}}{x_{N-2}} \\
& \times \cdots \matrixel{x_1}{e^{- \frac{i}{\hbar} \hat{H} \Delta t}}{x_0} \braket{x_0}{\psi(t_0)} \ ,
\end{split}
\end{equation}
where $\Delta t = (t_f - t_0)/N$. Then each small time propagator matrix element can be evaluated with the help of the momentum eigenket resolution of the identity (Eq. \ref{eq:QMpROI}). The key part of the calculation is evaluating matrix elements like these:
\begin{equation} \label{eq:SEmelcalc}
\begin{split}
\matrixel{x_{j+1}}{\hat{p^2}}{x_j} &= \frac{1}{2\pi \hbar} \int_{-\infty}^{\infty} dp \ \matrixel{x_{j+1}}{\hat{p^2}}{p} \braket{p}{x_j} \\
&=  \frac{1}{2\pi \hbar} \int_{-\infty}^{\infty} dp \ p^2 \braket{x_{j+1}}{p} \braket{p}{x_j} \\
&=  \frac{1}{2\pi \hbar} \int_{-\infty}^{\infty} dp \ p^2 e^{\frac{i}{\hbar} p (x_{j+1} - x_j)} \ .
\end{split}
\end{equation}
The rest of the calculation is fairly straightforward. First one arrives at the phase space path integral, which reads \cite{shankar2011}
\begin{equation}\label{eq:QMphase}
\begin{split}
\psi(x_0 \to x_f) &= \int \mathcal{D}[x(t)] \mathcal{D}[p(t)] \ \exp\left\{\frac{i}{\hbar} S[x, p] \right\} \\
S[x,p] &= \int_{t_0}^{t_f} dt \ p(t) \dot{x}(t) - \frac{p(t)^2}{2m}  - V(x(t)) \ .
\end{split}
\end{equation}
Then one can integrate out the momentum variables to arrive at the usual configuration space path integral
\begin{equation} \label{eq:QMconfig}
\begin{split}
\psi(x_0 \to x_f) &= \int \mathcal{D}[x(t)] \  \exp\left\{ \frac{i}{\hbar} S[x(t)] \right\}  \\
S[x] &= \int_{t_0}^{t_f} dt \ \frac{1}{2} m \dot{x}(t)^2  - V(x(t))
\end{split}
\end{equation}
which is over all paths satisfying $x(t_0) = x_0$ and $x(t_f) = x_f$. 

\subsection{Doi-Peliti path integral derivation}

The Doi-Peliti path integral is perhaps less familiar to most physicists, but the idea behind its derivation is the same as the idea behind the quantum mechanics derivation. 

Instead of the Schr\"{o}dinger equation (Eq. \ref{eq:schrodinger}), we begin with a master equation. Because thinking about general master equations involves juggling a lot of extra notational baggage, we will sketch the Doi-Peliti path integral derivation for a specific model: the birth-death process. Our list of chemical reactions reads
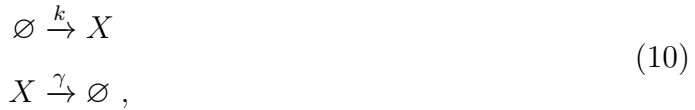
\begin{equation} \label{eq:rxns}
\begin{split}
\varnothing &\xrightarrow{k} X \\
X &\xrightarrow{\gamma} \varnothing \ ,
\end{split}
\end{equation}
where $k$ and $\gamma$ parameterize the rates of birth and death, respectively. The corresponding chemical master equation (CME) reads
\begin{equation} \label{eq:CME}
\begin{split}
\frac{\partial P(n, t)}{\partial t} =& \ k \left[ P(n-1,t) - P(n,t) \right] \\
&+ \gamma \left[ (n+1) P(n+1, t) - n P(n, t) \right] \ ,
\end{split}
\end{equation}
where $P(n, t)$ is the probability that the system has $n$ $X$ molecules at time $t$ (with $n \in \{0, 1, 2, ...\}$). 

In the spirit of Dirac's bra-ket formalism, we can rewrite the CME as an equation describing the time evolution of the so-called generating function $\ket{\psi}$, which is defined as the vector
\begin{equation} \label{eq:dpGF}
\ket{\psi} := \sum_{n = 0}^{\infty} P(n, t) \ket{n}
\end{equation}
in a Hilbert space spanned by the $\ket{n}$ vectors. In terms of this generating function, the CME reads
\begin{equation} \label{eq:dpEOM}
\begin{split}
\frac{\partial}{\partial t} \ket{\psi} &= \hat{H} \ket{\psi} \\
&= \left[ k (\hat{a}^+ - 1) + \gamma (\hat{a} - \hat{a}^+ \hat{a}) \right] \ket{\psi} \ ,
\end{split}
\end{equation}
where $\hat{H}$ is the Hamiltonian operator\footnote{Some older literature (Peliti's 1985 paper included) refers to this operator as the ``Liouvillian'' as opposed to the ``Hamiltonian''. We choose to call it the Hamiltonian both to emphasize the analogy with quantum mechanics, and to avoid notational collision when we talk about Lagrangians later.}, and the $\hat{a}$ and $\hat{a}^+$ are annihilation and creation operators that act linearly on the $\ket{n}$ kets according to
\begin{equation} \label{eq:annihilation}
\hat{a} \ket{n} = n \ket{n-1} 
\end{equation}
\begin{equation} \label{eq:creation}
\hat{a}^+ \ket{n} = \ket{n+1} \ .
\end{equation}
Just as above, this is formally solved by
\begin{equation} \label{eq:dpformalsln}
\ket{\psi(t_f)} = e^{\hat{H} (t_f - t_0)} \ket{\psi(t_0)}
\end{equation}
where $\ket{\psi(t_0)}$ is the system's state at initial time $t_0$, and $\ket{\psi(t_f)}$ is the system's state at a later time $t_f$. 

We would like to rewrite this by inserting many resolutions of the identity. In this case, the relevant resolution of identity is
\begin{equation} \label{eq:dpROI}
1 = \int_{-\infty}^{\infty} \int_{-\infty}^{\infty}  \frac{dz dz'}{2 \pi} \ \ket{z} \bra{- iz'} e^{- i z z'} \ ,
\end{equation}
corresponding to the (exclusive) inner product \cite{grassberger1980}
\begin{equation} \label{eq:exinnerprod}
\braket{n}{m} := n! \delta_{nm} \ ,
\end{equation}
with the so-called `coherent states' $\ket{z}$ defined via
\begin{equation} \label{eq:csdef}
\ket{z} = \sum_{n = 0}^{\infty} \frac{z^n}{n!} \ket{n} 
\end{equation}
for arbitrary $z \in \mathbb{C}$. What makes the coherent states useful is that they satisfy
\begin{equation} \label{eq:csprop}
\hat{a} \ket{z} = z \ket{z} \ .
\end{equation}
Moreover, since $\hat{a}$ and $\hat{a^+}$ are Hermitian conjugates of each other with respect to our inner product (Eq. \ref{eq:exinnerprod}), we have for any operator $\mathcal{O}$,
\begin{equation} \label{eq:dpmelexample}
\begin{split}
\matrixel{z_1}{\mathcal{O} }{z_2} &= \matrixel{z_1}{\sum_{n m} c_{nm} (\hat{a^+})^m (\hat{a})^n }{z_2} \\
&= \braket{z_1}{z_2} \sum_{n m} c_{nm} (z_1^*)^m (z_2)^n \\
&= e^{z_1^* z_2} \sum_{n m} c_{nm} (z_1^*)^m (z_2)^n \ .
\end{split}
\end{equation}
Upon inserting the coherent state resolution of the identity many times into Eq. \ref{eq:dpformalsln}, we obtain
\begin{equation} \label{eq:dpsandwich}
\begin{split}
\ket{\psi(t_f)} =& e^{\hat{H} \Delta t} e^{\hat{H} \Delta t} \cdots e^{\hat{H} \Delta t} \ket{\psi(t_0)} \\
=& \int \frac{dz_0 dz_0'}{2\pi} \cdots \frac{dz_N dz_N'}{2 \pi} \ket{z_N}  \\
& \times \matrixel{- i z_N'}{e^{ \hat{H} \Delta t}}{z_{N-1}} \matrixel{- i z_{N-1}'}{e^{ \hat{H} \Delta t}}{z_{N-2}} \\
& \times \cdots \matrixel{- i z_1'}{e^{\hat{H} \Delta t}}{z_0} \braket{- i z_0'}{\psi(t_0)} \ ,
\end{split}
\end{equation}
where $\Delta t = (t_f - t_0)/N$. From here, the coherent state property (Eq. \ref{eq:csprop}) can be exploited to finish the calculation and arrive at the Doi-Peliti path integral. 

It should probably be noted, for clarity's sake, that the Doi-Peliti path integral is qualitatively somewhat different from the quantum mechanics one described in the previous subsection. In quantum mechanics and quantum field theory, there are two broad classes of path integrals: one involving integration over many paths through configuration space or phase space (i.e. what we just described), and another involving integration over \textit{coherent states} \cite{klauder1960, bargmann1961, glauber1963, klauder2001}. Because the Doi-Peliti construction involves integrating over coherent states, it is an example of the second kind of path integral. It is also possible, although less popular, to write down a path integral which corresponds more directly to the usual quantum mechanics one. This one involves summing over all of the possible paths through discrete state space, so one has an infinite number of sums instead of an infinite number of integrals \cite{weber_master_2017}. 

%The latter are commonly used in quantum optics, or to describe Ising-type systems. 
%rather than paths through the system's discrete state space, 

%=======================================================================

\section{Variable types and model assumptions}
\label{sec:vtypes}

%\subsection{Assumptions and possible variable types}

%General purpose stochastic processes refs: 

In order to present a relatively simple-looking derivation in the following section, we will focus on stochastic systems defined by a one-dimensional Ito-interpreted \cite{ito1944} stochastic differential equation (SDE)
\begin{equation}\label{eq:1Dsde}
\dot{x} = f(x) + g(x) \eta(t) \ ,
\end{equation}
where $\eta(t)$ is a Gaussian white noise term, and $f$ and $g$ have no explicit time-dependence. This equation might represent how the concentration level of some molecule changes stochastically with time, or how the position of a Brownian particle changes as it is bombarded by other particles. We describe how to treat much more general systems (time-dependent $f$ and $g$, multiple variables, and so on) in Sec. \ref{sec:modifications}. 

The probability $P(x, t)$ that our system is in state $x$ at time $t$ (given some initial condition $P_0(x)$) satisfies the Fokker-Planck equation \cite{vankampen2007, gardiner2009}
\begin{equation}\label{eq:1DItoFP}
\frac{\partial P(x,t)}{\partial t} = - \frac{\partial}{\partial x} \left[ f(x) P(x,t)  \right] + \frac{1}{2} \frac{\partial^2}{\partial x^2} \left[ g(x)^2 P(x,t)  \right] \ .
\end{equation}
Because---at least in principle---any question about a system described by Eq. \ref{eq:1Dsde} can be answered using the time-dependent probability density $P(x, t)$, our goal is to construct a formal solution to Eq. \ref{eq:1DItoFP}. 
%For simplicity, we will in this section focus on the Fokker-Planck equation associated with a one-dimensional Ito-interpreted stochastic differential equation (SDE)

%Although $f$ and $g$ may be time-dependent in general, they are usually not for many applications of interest (like gene regulation models). In what follows, we will assume they are time-independent, and refer the reader to Sec. \ref{sec:modifications} for how to generalize our result to time-dependent $f$ and $g$.\\

%we will allow them to be time-dependent, since it will not affect our calculation (although we will not explicitly write the time-dependence in what follows to ease notation). \\

Now we must comment on the domain of our variable $x$. We will assume that $x \in [0, \infty)$, since we are motivated by gene regulation literature, and it is clear that negative concentrations are generally not realistic. We also assume an Ito-interpreted SDE in this section since it is the correct interpretation for chemical Langevin equations\footnote{See Eq. 22 of Gillespie \cite{gillespie2000} and note that it takes the form of an Euler-Maruyama time step.}. 

However, there are other possible choices of variable domain. It is often assumed that $x \in (-\infty, \infty)$, especially when the Fokker-Planck equation is used to describe diffusion processes. It is also logically possible that $x$ takes values in an interval, i.e. $x \in [a, b]$ for some real constants $a$ and $b$ with $a \leq b$. 

It turns out that the formalism we describe must be modified slightly when considering these different types of variables (see Sec. \ref{sec:modifications}). Hence, for convenience, we will introduce the following terminology:
\begin{itemize}
\item \textbf{diffusion-type variables} \\take values in $(-\infty, \infty)$
\item \textbf{concentration-type variables} \\take values in $[0, \infty)$
\item \textbf{compact-type variables} \\take values in $[a, b]$ for some $a \leq b$
\end{itemize}
In each of these cases, the appropriate `momentum' eigenstates, along with their associated resolution of the identity, will change. 

If a variable has a domain of the form $[a, \infty)$ for some $a \neq 0$, or one of the form $(-\infty, a]$, it can be rewritten as a concentration-type variable (in the sense described above) by a trivial change of variables. For this reason, we will consider variables with such domains as concentration-type.

\begin{table*}[h]
%\centering
\begin{tabular}{ |p{3.6cm}||p{1.6cm}|p{3.9cm}|p{7cm}|  }
 \hline
 \multicolumn{4}{|c|}{Summary of different variable types} \\
 \hline
 Variable type  & Domain & Usual interpretation & Representative examples \\
 \hline \hline
 Concentration-type   & $[0, \infty)$ & Ito \cite{gillespie2000}  & chemical kinetics, gene regulation, population dynamics \\
 Diffusion-type &   $(-\infty, \infty)$  &  Stratonovich \cite{gardiner2009}\cite{vankampen2007}  & diffusion, statistical mechanics, quantum mechanics \\
 Compact-type & $[a, b]$ &  depends on context  & confined diffusion, enzyme kinetics\\
 \hline
\end{tabular}
\caption{Summary of the different variable types one might consider in a SDE model. }
\label{tab:variablesummary}
\end{table*}

%=======================================================================

\section{The one variable derivation}
\label{sec:1Dderivation}

Sec. \ref{sec:review} can be summarized as follows. In quantum mechanics and Doi-Peliti field theory, we can find time-dependent solutions to our equations of interest (the Schr\"{o}dinger equation and the master equation, respectively) by constructing a formal path integral solution. To construct this solution, we
\begin{enumerate}
\item Reframe the original problem in terms of vectors in some Hilbert space, and linear operators that act on those vectors.
\item Construct a formal solution of the reframed equation using many resolutions of the identity.
\item Exploit the relationship between the operators in the Hamiltonian and the inserted states to evaluate matrix elements and write down the final path integral.
\end{enumerate}
In this section, we will show that exactly the same strategy can be used to solve the Fokker-Planck equation, Eq. \ref{eq:1DItoFP}.

% Note that the prescription for compact-type variables reduces to the diffusion-type variable prescription in the L -> infty limit. 
% Are there multiple kinds of compact-type variables? Should there be a prescription that reduces to the concentration-type prescription in the L -> infty limit? 

% ===============================================================

\subsection{Reframing of problem in Hilbert space}

\subsubsection{States and inner product}

Our first step, \textit{a la} Peliti's derivation, is to modify the problem we are trying to solve. For $P(x, t)$ corresponding to the Ito-interpreted concentration-type variable problem described in Sec. \ref{sec:vtypes}, consider the generating function\footnote{The technique of rephrasing one's problem in terms of a generating function is common in the study of stochastic processes governed by things like the CME and SDEs, as well as in probability theory more generally. For a charming book on generating function approaches to problems in combinatorics, number theory, and probability, see Wilf \cite{wilf2006}.} defined by
\begin{equation} \label{eq:gf}
\ket{\psi(t)} := \int_0^{\infty} dx \ P(x,t) \ket{x} \ ,
\end{equation}
which lives in a Hilbert space spanned by the $\ket{x}$ vectors. Define an arbitrary state in this space via
\begin{equation} \label{eq:arbket}
\ket{\phi} := \int_0^{\infty} dx \ c(x) \ket{x} 
\end{equation}
where $c(x)$ is allowed to be complex-valued. For the state we are most interested in (Eq. \ref{eq:gf}), $c(x)$ corresponds to the state space probability distribution $P(x, t)$; however, we will also consider states (for example, see the diffusion-type momentum eigenkets defined in Sec. \ref{sec:modifications}) where $c(x)$ does not necessarily correspond to anything physically or biologically meaningful. 
 
Define the corresponding bra
\begin{equation} \label{eq:arbbra}
\bra{\phi} := \int_0^{\infty} dx \ c^*(x) \bra{x}  \ .
\end{equation}
Define a scalar product on the $\ket{x}$ vectors via
\begin{equation} \label{eq:scalarproddef1}
\braket{x}{x'} := \delta(x - x')  \ ,
\end{equation}
and extend it via linearity to arbitrary states so that
\begin{equation} \label{eq:scalarproddef2}
\braket{\phi_2}{\phi_1} := \int_0^{\infty} dx \ c_2^*(x) c_1(x) \ .
\end{equation}

Now we will try to solve for the time evolution of the generating function $\ket{\psi(t)}$, and recover the time evolution of the probability density described by Eq. \ref{eq:1DItoFP} as a consequence. 

For an operator $\hat{\mathcal{O}}$, the meaning of 
\begin{equation} \label{eq:matrixel}
\matrixel{\phi_2}{\hat{\mathcal{O}}}{\phi_1}
\end{equation}
is that $\hat{\mathcal{O}}$ always acts to the right. Unlike in quantum mechanics, the operators we will consider (like the momentum operator for concentration-type variables) are in general not Hermitian with respect to the inner product given by Eq. \ref{eq:scalarproddef2}.

\subsubsection{Operators}

By analogy with quantum mechanics, we will define two important operators: the state operator $\hat{x}$, and the momentum operator $\hat{p}$. Define the state operator $\hat{x}$ by its action on $\ket{x}$ vectors:
\begin{equation} \label{eq:xopdef}
\hat{x} \ket{y} := y \ket{y} \ ,
\end{equation}
where as usual we use the convention that an $\ket{x}$ vector is labeled by its eigenvalue. Similarly, associate with any function $h(x)$ an operator $h(\hat{x})$ that acts according to
\begin{equation} \label{eq:xopfunctiondef}
h(\hat{x}) \ket{y} := h(y) \ket{y} \ .
\end{equation}

Define the `momentum' operator $\hat{p}$ by
\begin{equation} \label{eq:popdef}
\hat{p} \ket{\psi} = - \int_{0}^{\infty} dx \ \frac{\partial c}{\partial x} \ket{x} \ ,
\end{equation}
\textit{a la} quantum mechanics. One way to rationalize this choice is to try writing an infinitesimal translation operator in the form $\exp(\hat{p} \Delta x)$. Like in quantum mechanics, these operators satisfy a canonical commutation relation $[\hat{x} , \hat{p}] = 1$.

\subsubsection{Momentum eigenkets}
\label{subsubsec:peigenket}
The eigenkets of momentum, i.e. states $\ket{p'}$ which satisfy
\begin{equation} \label{eq:popdef1}
\hat{p} \ket{p'} = p' \ket{p'} \ ,
\end{equation}
can be shown to be equal to 
\begin{equation} \label{eq:popdef2}
\ket{p'} = \int_0^{\infty} dx \ e^{- p' x} \ket{x} 
\end{equation}
up to a multiplicative constant.\footnote{Choosing a constant other than $1$ just means a different constant will appear in front of the momentum eigenket resolution of the identity, Eq. \ref{eq:ROIp}.} For later use, we will record that
\begin{equation} \label{eq:xpinnerprod}
\braket{x}{p} = e^{- px} \ , \ \braket{p}{x} = e^{- p^* x} \ .
\end{equation}

Interestingly, thinking about the `momentum-space' probability density $P(p, t)$ for real $p$ corresponds to taking a Laplace transform:
\begin{equation} \label{eq:pspaceprob}
\begin{split}
P(p, t) &:= \braket{p}{\psi} \\
&= \int_0^{\infty} dx \ \braket{p}{x} \braket{x}{\psi} \\
&= \int_0^{\infty} dx \ e^{- p x} P(x, t) \ .
\end{split}
\end{equation}
Compare this with quantum mechanics, where moving from position space to momentum space corresponds to taking a Fourier transform. This change is one of the significant consequences of considering a concentration-type variable instead of a diffusion-type variable (for which we will see moving to momentum space is a Fourier transform in Sec. \ref{sec:modifications}). 

Importantly, thinking backwards, we realize that we can \textit{define} the momentum operator $\hat{p}$ and the associated momentum eigenkets such that the momentum space probability density $P(p, t)$ corresponds to taking any integral transform we like. Choosing $P(p, t)$ to be the Laplace transform is particularly appropriate here, because $x \in [0, \infty)$ for concentration-type variables; however, for other variable types, different transforms would be more appropriate. See Sec. \ref{sec:modifications} for more discussion of this idea.

Finally, note that $p'$ is in principle allowed to be whatever---even a complex number. This will be important in the next subsection.

\subsubsection{Resolutions of the identity} 

Since, for an arbitrary state $\ket{\phi}$, 
\begin{equation}
\begin{split}
& \int_0^{\infty} dx \ \ket{x} \braket{x}{\phi} \\
=& \int_0^{\infty} dx \int_0^{\infty} dx' \ \ket{x} c(x') \braket{x}{x'} \\
=& \int_0^{\infty} dx \int_0^{\infty} dx' \ \ket{x} c(x') \delta(x - x') \\
=& \int_0^{\infty} dx \ c(x) \ket{x} \\
=& \ket{\phi} \ ,
\end{split}
\end{equation}
we have a resolution of the identity
\begin{equation} \label{eq:ROIx}
1 = \int_0^{\infty} dx \ \ket{x} \bra{x} \ .
\end{equation}
Similarly, one can show that we have another resolution of the identity
\begin{equation} \label{eq:ROIp}
1 = \frac{1}{2\pi i} \int_{-i \infty}^{i \infty} dp \ \ket{p} \bra{p} 
\end{equation}
in terms of momentum eigenkets. 

Following our comment at the end of the previous section, notice that this formula looks much like the Bromwich integral/Mellin's inverse formula \cite{cohen2007}, the formula for an inverse Laplace transform. Indeed, going from momentum space to position space exactly corresponds to taking an inverse Laplace transform.

\subsubsection{Hamiltonian and equation of motion}

Given the Fokker-Planck equation and our generating function given by Eq. \ref{eq:gf}, we have that
\begin{equation} 
\begin{split}
\frac{\partial \ket{\psi(t)}}{\partial t} =&  \int_0^{\infty} dx \ \frac{\partial P(x,t)}{\partial t} \ket{x} \\
=&  \int_0^{\infty} dx \ \left( - \frac{\partial}{\partial x} \left[ f(x) P(x,t)  \right] \right. \\
& \left. + \frac{1}{2} \frac{\partial^2}{\partial x^2} \left[ g(x)^2 P(x,t)  \right] \right) \ket{x} \\
=& \left( \hat{p} f(\hat{x}) + \frac{1}{2} \hat{p}^2 g(\hat{x})^2  \right) \int_0^{\infty} dx \ P(x,t) \ket{x} \\
=& \hat{H} \ket{\psi} \ ,
\end{split}
\end{equation}
where we define the Hamiltonian operator $\hat{H}$ as
\begin{equation} \label{eq:Hopdef}
\hat{H} := \hat{p} f(\hat{x}) + \frac{1}{2} \hat{p}^2 g(\hat{x})^2 \ .
\end{equation}
Similar choices of Hamiltonian \cite{ZinnJustin2002, atland2010, parisi1988, kamenev2011, weber_master_2017, kamenev2008, escudero2009, parker2009, koide2017, jizba2008, graham1977, mizrahi1981} have been made before, although conventions differ depending on how $\hat{p}$ is defined, and on whether certain overall constants are considered part of the Hamiltonian or separate from it. 

The generating function's equation of motion is
\begin{equation} \label{eq:gfEOM}
\frac{\partial \ket{\psi(t)}}{\partial t}  = \hat{H} \ket{\psi} \ .
\end{equation}
This is clearly analogous to the bra-ket notation Schr\"{o}dinger equation (Eq. \ref{eq:schrodingerbraket}). It is this equation, instead of Eq. \ref{eq:1DItoFP}, that we will solve. 

%In other words, we have another resolution of the identity

%\subsubsection{Momentum eigenket resolution of the identity}

%\subsubsection{Special matrix elements}
%
%In general, our operators are not expected to be Hermitian---indeed, $\hat{p}$ is not, at least with respect to the inner product \ref{eq:scalarproddef2} that we defined. \\
%
%It is easy to show that
%\begin{equation} \label{eq:xpinnerprod}
%\braket{x}{p} = e^{- px} \ , \ \braket{p}{x} = e^{- p^* x} \ .
%\end{equation}

% =====================================================================

\subsection{Formal solution of reframed problem}

\subsubsection{Naive formal solution}

Eq. \ref{eq:gfEOM} has the usual formal solution
\begin{equation} \label{eq:EOMformalsolution}
\ket{\psi(t_f)} = e^{\hat{H} (t_f - t_0)} \ket{\psi(t_0)} \ .
\end{equation}
We can if we like define the time evolution operator $\hat{U}(t_f, t_0)$ as
\begin{equation} \label{eq:Uopdef}
\hat{U}(t_f, t_0) := e^{\hat{H} (t_f - t_0)} \ ,
\end{equation}
and talk about Eq. \ref{eq:EOMformalsolution} in terms of it. For Hamiltonians $\hat{H}$ without explicit time-dependence, we may also write $\hat{U}(t_f - t_0)$, since only the time difference matters in that case.

%\subsubsection{Decomposition into many time steps}

Take $t_j = t_0 + j \Delta t$, where $\Delta t = \frac{t_f - t_0}{N}$ is very small, and decompose the time evolution operator into $N$ small time steps:
\begin{equation}
\begin{split}
& \ket{\psi(t_f)} \\
&= \hat{U}(t_f - t_{N-1}) \hat{U}(t_{N-1} - t_{N-2}) \cdots \hat{U}(t_1 - t_0)  \ket{\psi(t_0)} \\
&= \hat{U}(\Delta t) \hat{U}(\Delta t) \cdots \hat{U}(\Delta t)  \ket{\psi(t_0)} \ .
\end{split}
\end{equation}
Now insert $N+1$ resolutions of the identity:
\begin{equation}
\begin{split}
& \ket{\psi(t_f)} \\
&= \int_0^{\infty} dx_0 \int_0^{\infty} dx_1 \cdots \int_0^{\infty} dx_N \ \ket{x_N} \\
& \times \matrixel{x_N}{\hat{U}(\Delta t)}{x_{N-1}}  \cdots \matrixel{x_1}{\hat{U}(\Delta t)}{x_0}  \braket{x_0}{\psi(t_0)} \\
&= \int \left[ \prod_{j = 0}^N dx_j \right] \ket{x_N} \ \left[ \prod_{j = 1}^N \matrixel{x_j}{\hat{U}(\Delta t)}{x_{j-1}} \right] P_0(x_0) 
\end{split}
\end{equation}
where we have used the fact that $\braket{x}{\psi(t)} = P(x,t)$. 

Since all other probabilities may be written in terms of it, we may as well specialize to the transition probability $P(x_f, t_f; x_0, t_0)$. To do this, choose $P_0(x) = \delta(x - x_0)$ and note that $P(x_f, t_f; x_0, t_0) = \braket{x_f}{\psi(t_f)}$. Then we have
\begin{equation} \label{eq:tprobexpansion}
\begin{split}
& P(x_f, t_f; x_0, t_0) \\
=& \int_0^{\infty} dx_1 \cdots \int_0^{\infty} dx_{N-1} \matrixel{x_f}{\hat{U}(\Delta t)}{x_{N-1}} \\
& \cdots \matrixel{x_1}{\hat{U}(\Delta t)}{x_0}  \ .
\end{split}
\end{equation}
This expression is analogous to the one for the propagator (or transition `amplitude') from ordinary quantum mechanics \cite{feynman2010, schulman1981, kleinert2009}. In any case, all that remains is to evaluate the matrix elements of the infinitesimal time evolution operators. 

% ======================================================================

\subsection{Evaluating matrix elements and finishing the calculation}

\subsubsection{Evaluating time evolution operator matrix elements}

We would like to evaluate the matrix element 
\begin{equation}
\matrixel{x_j}{\hat{U}(\Delta t)}{x_{j-1}} \ .
\end{equation}
Since $\Delta t$ is very small, we have (to first order in $\Delta t)$
\begin{equation}
\hat{U}(\Delta t) = 1 + \hat{H} \Delta t \ .
\end{equation}
By linearity, we can write
\begin{equation}
\begin{split}
\matrixel{x_j}{\hat{U}(\Delta t)}{x_{j-1}} &= \braket{x_j}{x_{j-1}} + \matrixel{x_j}{\hat{H}}{x_{j-1}} \Delta t \\
&= \delta(x_j - x_{j-1}) + \matrixel{x_j}{\hat{H}}{x_{j-1}} \Delta t \ .
\end{split}
\end{equation}
Again by linearity, we have
\begin{equation}
\begin{split}
& \matrixel{x_j}{\hat{H}}{x_{j-1}} \\
&= \matrixel{x_j}{\hat{p} f(\hat{x}) + \frac{1}{2} \hat{p}^2 g(\hat{x})^2}{x_{j-1}} \\
&= \matrixel{x_j}{\hat{p} f(\hat{x})}{x_{j-1}} + \frac{1}{2} \matrixel{x_j}{\hat{p}^2 g(\hat{x})^2}{x_{j-1}} \\
&= f(x_{j-1}) \matrixel{x_j}{\hat{p}}{x_{j-1}} + \frac{1}{2} g(x_{j-1})^2 \matrixel{x_j}{\hat{p}^2}{x_{j-1}} \ .
\end{split}
\end{equation}
Here we must use our momentum eigenket resolution of the identity. Note,
\begin{equation}
\begin{split}
\matrixel{x_j}{\hat{p}}{x_{j-1}} &= \frac{1}{2\pi i} \int_{-i \infty}^{i \infty} dp \ \matrixel{x_j}{\hat{p}}{p} \braket{p}{x_{j-1}} \\
&= \frac{1}{2\pi i} \int_{-i \infty}^{i \infty} dp \ p \braket{x_j}{p} e^{- p^* x_{j-1}} \\
&= \frac{1}{2\pi i} \int_{-i \infty}^{i \infty} dp \ p e^{- p x_j} e^{- p^* x_{j-1}} \ .
\end{split}
\end{equation}
Perform a change of variables $u = - ip$. Then we have
\begin{equation}
\matrixel{x_j}{\hat{p}}{x_{j-1}} = \int_{- \infty}^{ \infty} \frac{du}{2\pi} \ i u e^{- i u (x_j - x_{j-1})}  \ .
\end{equation}
Similarly,
\begin{equation}
\matrixel{x_j}{\hat{p}^2}{x_{j-1}} = \int_{- \infty}^{ \infty} \frac{du}{2\pi} \ (-u^2) e^{- i u (x_j - x_{j-1})} \ .
\end{equation}
Combining these results, we find that $\matrixel{x_j}{\hat{H}}{x_{j-1}}$ is
\begin{equation} 
\int_{- \infty}^{ \infty} \frac{du}{2\pi} e^{- i u (x_j - x_{j-1})} \left[ i u f(x_{j-1}) - \frac{1}{2} u^2 g(x_{j-1})^2  \right] \ . 
\end{equation}

Now we can write
\begin{equation} 
\begin{split}
& \matrixel{x_j}{\hat{U}(\Delta t)} {x_{j-1}} \\ =& \delta(x_j - x_{j-1}) + \matrixel{x_j}{\hat{H}}{x_{j-1}} \Delta t \\
=& \int_{-\infty}^{\infty} \frac{du}{2\pi} \ e^{- i u (x_j - x_{j-1})} \\
&\times \left( 1 +  \left[ i u f(x_{j-1}) - \frac{1}{2} u^2 g(x_{j-1})^2  \right] \Delta t  \right)  \\
=& \int_{-\infty}^{\infty} \frac{du}{2\pi} \ e^{- i u (x_j - x_{j-1})} e^{i u f(x_{j-1}) \Delta t - \frac{1}{2} u^2 g(x_{j-1})^2 \Delta t  }  \\
\end{split}
\end{equation}
where the last equality is justified from $\Delta t$ being infinitesimally small (so the exponential is equal to its first order Taylor expansion). Written in a more suggestive form, this result reads
\begin{equation} \label{eq:Umatrixel}
\begin{split}
& \matrixel{x_j}{\hat{U}(\Delta t)}{x_{j-1}} \\
=& \int_{-\infty}^{\infty} \frac{dp_j}{2 \pi} \ \exp\left\{- \left[ i p_j \left( \frac{x_j - x_{j-1}}{\Delta t} - f(x_{j-1}) \right) \right. \right.   \\
 &+ \left. \left.\frac{1}{2} {p_j}^2 g(x_{j-1})^2   \right] \Delta t \right\} \ ,
\end{split}
\end{equation}
where we have relabeled $u$ as $p_j$ (since there will ultimately be $N$ of these dummy variables together in the same expression).

\subsubsection{Computing the propagator by combining matrix elements}

Using Eq. \ref{eq:Umatrixel}, we have
\begin{widetext} 
\begin{equation} \label{eq:discreteMSRJD}
P(x_f, t_f; x_0, t_0) = \lim_{N \to \infty} \int \frac{dp_N}{2\pi}\prod_{j = 1}^{N-1} \ \frac{dx_j dp_j}{2\pi} \ \exp{- \sum_{j = 1}^N \left[ i p_j \left( \frac{x_j - x_{j-1}}{\Delta t} - f(x_{j-1}) \right) + \frac{1}{2} {p_j}^2 g(x_{j-1})^2 \right] \Delta t} 
\end{equation}
\end{widetext}

Although some practical calculations are more clearly carried out by doing the finite number of integrals first, and then taking $N$ to infinity, this result is often schematically written as
\begin{equation} \label{eq:schematicMSRJD}
P(x_f, t_f; x_0, t_0) = \int \mathcal{D}[x(t)] \mathcal{D}[p(t)] \ \exp\left\{- S[x, p] \right\} \ ,
\end{equation}
where the factors of $2\pi$ have been absorbed into the measure, and the integration is done over all paths $x(t)$ and $p(t)$ with $x(t_0) = x_0$ and $x(t_f) = x_f$. The action $S$ is the functional defined by
\begin{equation}\label{eq:MSRJDaction}
S[x, p] :=  \int_{t_0}^{t_f} dt \ i p(t) \left[ \dot{x}(t) - f(x(t)) \right] + \frac{1}{2} {p(t)}^2 g(x(t))^2 
\end{equation}
and the corresponding Lagrangian is
\begin{equation}\label{eq:MSRJDLagrangian}
L :=  i p \left[ \dot{x} - f(x) \right] + \frac{1}{2} {p}^2 g(x)^2 \ .
\end{equation}

\subsection{From the MSRJD to Onsager-Machlup path integral}

The path integral we just derived, whose action is given by Eq. \ref{eq:MSRJDaction}, is called \cite{cugliandolo2017, cugliandolo2019} the MSRJD functional or path integral (after Martin-Siggia-Rose \cite{msr1973}, Janssen \cite{janssen1976}, and De Dominicis \cite{dd1976}, although Peliti's \cite{ddpeliti1978} name is also sometimes included \cite{hertz2016}). For some purposes (like calculating correlation functions \cite{chow_path_2015}), it is a useful tool. However, the presence of an additional path variable $p(t)$ that contributes to the action can be inconvenient, especially when we are only interested in the `least action' path. We would like some way to remove it, so that the integral is \textit{only} over our physically/biologically relevant variable, and not over any other dummy variables. 

Moving from the MSRJD path integral to the Onsager-Machlup path integral is very much like moving from the phase space path integral to the configuration space path integral in quantum mechanics (c.f. Eq. \ref{eq:QMphase} and Eq. \ref{eq:QMconfig}). 

As an aside, the MSRJD and phase space path integral actions, while generally different, correspond in some cases. For example, the action for a charged particle in a certain magnetic field can correspond exactly to a diffusion-type variable with additive noise \cite{wiegel1981, graham1977, hie1985}.  

The $N$ integrals over the response variables are all Gaussian, and so can be easily carried out, yielding 
\begin{widetext} 
\begin{equation} \label{eq:discreteOM}
\begin{split}
P(x_f, t_f; x_0, t_0) &= \lim_{N \to \infty} \int \left[ \prod_{j = 1}^{N-1} dx_j \right] \prod_{j = 1}^{N} \ \frac{1}{2\pi} \ \sqrt{\frac{2\pi}{g(x_{j-1})^2 \Delta t}} \exp\left\{ - \frac{\left[  \frac{x_j - x_{j-1}}{\Delta t} - f(x_{j-1})\right]^2}{2 g(x_{j-1})^2} \Delta t   \right\}  \\
&= \lim_{N \to \infty} \int \left[ \prod_{j = 1}^{N-1} dx_j \right] \left[  \prod_{j = 1}^{N} \frac{1}{\sqrt{2 \pi g(x_{j-1})^2 \Delta t}} \right] \exp\left\{ - \sum_{j = 1}^N \frac{\left[  \frac{x_j - x_{j-1}}{\Delta t} - f(x_{j-1})\right]^2}{2 g(x_{j-1})^2} \Delta t   \right\}  \ .
\end{split}
\end{equation}
\end{widetext}
This is a sort of Onsager-Machlup path integral (originally defined for additive noise \cite{OM1953pt1, OM1953pt2}, and later generalized to allow for state-dependent noise \cite{graham1977, arnold2000}). Schematically, it can be written as
\begin{equation} \label{eq:schematicOM}
P(x_f, t_f; x_0, t_0) =  \int \mathcal{D}[x(t)] \ \exp\left\{ - S[x(t)] \right\} 
\end{equation}
where the noise-dependent prefactors have been absorbed into the measure definition, and the integration is over all paths $x(t)$ with $x(t_0) = x_0$ and $x(t_f) = x_f$. Reading off the argument of the exponential in Eq. \ref{eq:discreteOM}, the action $S$ in the continuum limit is
\begin{equation} \label{eq:OMaction}
S[x(t)] = \int_{t_0}^{t_f} dt \ \frac{\left[  \dot{x}(t) - f(x(t))\right]^2}{2 g(x(t))^2}  \ .
\end{equation}
This corresponds to a Lagrangian
\begin{equation} \label{eq:OMLagrangian}
L = \frac{\left[  \dot{x}(t) - f(x(t))\right]^2}{2 g(x(t))^2}  \ ,
\end{equation}
which can be used (via, say, the Euler-Lagrange equations) to find the most likely transition path between any initial state and final state. This Lagrangian can also be obtained directly from the MSRJD Lagrangian (Eq. \ref{eq:MSRJDLagrangian}) by using the Euler-Lagrange equations to write $p$ in terms of $x$.

%=======================================================================

\section{Modifications and extensions to the formalism}
\label{sec:modifications}

An appealing feature of our approach is that can be easily modified and extended to consider alternative problem types and formulations. Here we discuss a few of these.

\subsection{Diffusion and compact-type variables} \label{ssec:diffandcompact}

Why did we choose Eq. \ref{eq:popdef} as the prescription for our concentration-type momentum operator $\hat{p}$, and Eq. \ref{eq:popdef2} as the prescription for our concentration-type momentum eigenkets? We could have easily chosen the momentum operator to be $\partial/\partial x$ with some constant other than $-1$.

In one view, the observation that going to momentum space corresponds to taking a Laplace transform (see Sec. \ref{subsubsec:peigenket}) is actually the \textit{motivation} for those earlier choices; we chose our momentum conventions that way because the Laplace transformed problem is the `natural' conjugate problem for dynamics defined on $[0, \infty)$. Following this idea, we may decide the following: a Fourier transformed problem is the `natural' conjugate problem for dynamics defined on $(-\infty, \infty)$, and moving to Fourier coefficients is the `natural' conjugate problem for dynamics defined on an interval $[a, b]$. 

With this in mind, we define the momentum operator and associated eigenkets in the following way for diffusion-type variables:
\begin{equation} \label{eq:difftypep}
\begin{split}
\hat{p} &:= - i \frac{\partial}{\partial x} \\
\ket{p} &:= \int_{-\infty}^{\infty} dx \ e^{i p x} \ket{x} \\
1 &= \frac{1}{2\pi} \int_{-\infty}^{\infty} dp \ \ket{p} \bra{p} \ .
\end{split}
\end{equation}

The $i$ appears in this case not because we believe that diffusion-like processes are mysterious and different, but because we want considering the momentum-space probability density to be like taking a Fourier transform: 
\begin{equation} \label{eq:difftypefourier}
\begin{split}
P(p, t) &:= \braket{p}{\psi} \\
&= \int_{-\infty}^{\infty} dx \ \braket{p}{x} \braket{x}{\psi} \\
&= \int_{-\infty}^{\infty} dx \ e^{- i p x} P(x, t) \ .
\end{split}
\end{equation}

The final result for diffusion-type variables is exactly the same as for concentration-type variables (Eq. \ref{eq:discreteMSRJD}, Eq. \ref{eq:discreteOM}), but with all of the $x_j$ integrals from $-\infty$ to $\infty$ instead of from $0$ to $\infty$.

For compact-type variables, we define
\begin{equation} \label{eq:compacttypep}
\begin{split}
\hat{p} &:= - i \frac{\partial}{\partial x} \\
\ket{p} &:= \int_{a}^{b} dx \ e^{i p x} \ket{x} \\ 1 &= \frac{1}{L} \sum_{n = -\infty}^{\infty} \ket{p_n} \bra{p_n} \ \hspace{0.1in} \ , \ p_n = 2 \pi n/L  
\end{split}
\end{equation}
where $L := b - a$. Note that we must sum over certain eigenkets in the corresponding resolution of the identity, instead of integrating over them as usual. 

Moving to momentum space corresponds to moving to Fourier coefficients, except for a factor of $1/L$:
\begin{equation} \label{eq:compacttypefourier}
\begin{split}
P(p, t) &:= \braket{p}{\psi} \\
&= \int_{a}^{b} dx \ \braket{p}{x} \braket{x}{\psi} \\
&= \int_{a}^{b} dx \ e^{- i p x} P(x, t) \ .
\end{split}
\end{equation}

As explained before, putting the factor of $1/L$ in the resolution of the identity instead of the momentum eigenkets is just convention. The result for compact-type variables is
\begin{widetext} 
\begin{equation} \label{eq:discreteMSRJDcompact}
\begin{split}
P = \lim_{N \to \infty} \frac{1}{L^N}  \sum_{n_1 = - \infty}^{\infty} \cdots \sum_{n_N = -\infty}^{\infty} \int_a^b dx_1 \cdots \int_a^b dx_{N-1}   \exp{- \sum_{j = 1}^N \left[ i p_j \left( \frac{x_j - x_{j-1}}{\Delta t} - f(x_{j-1}) \right) + \frac{1}{2} {p_j}^2 g(x_{j-1})^2 \right] \Delta t}
\end{split}
\end{equation}
\end{widetext}
with $p_j = \frac{2 \pi n_j}{L}$ for all $j = 1, ..., N$. There is generally not a corresponding Onsager-Machlup form, because the sums over the $p_j$ variables are not tractable (unless one really likes working with theta functions).

For completeness' sake, we note that these are not the only choices for these variable types; in principle, one may choose momentum operators and eigenkets so that the associated momentum space problem corresponds to any transform one likes (although there is no guarantee that the resulting path integral will have a nice-looking expression).\\

\begin{table*}[h]
\centering
\begin{tabular}{ |p{3.6cm}||p{1.8cm}|p{2cm}|p{3cm}|p{3cm}|p{3cm}|  }
 \hline
 \multicolumn{6}{|c|}{Formalism modifications for different variable types} \\
 \hline
 Variable type  & Domain & $\hat{p}$ operator & $\hat{p}$ eigenket & R.O.I. & Corresponds to \\
 \hline \hline
 Concentration-type   & $[0, \infty)$ & $- \frac{\partial}{\partial x}$ & $\int_0^{\infty} dx \ e^{- p x} \ket{x}$ &   $\frac{1}{2\pi i} \int_{-i \infty}^{i \infty} dp \ \ket{p} \bra{p}$ & Laplace transform \\
 Diffusion-type &   $(-\infty, \infty)$  & $- i \frac{\partial}{\partial x}$   & $\int_{-\infty}^{\infty} dx \ e^{i p x} \ket{x}$  & $\frac{1}{2\pi} \int_{-\infty}^{\infty} dp \ \ket{p} \bra{p}$ & Fourier transform \\
 Compact-type & $[a, b]$ & $- i \frac{\partial}{\partial x}$ & $\int_{a}^{b} dx \ e^{i p x} \ket{x}$ & $\frac{1}{L} \sum_{n = -\infty}^{\infty} \ket{p} \bra{p}$ \ $p = 2 \pi n/L$ & Fourier series \\
 \hline
\end{tabular}
\caption{Summary of how our formalism changes for the different types of variables introduced in Sec. \ref{sec:vtypes}.}
\label{tab:formalismchanges1D}
\end{table*}

\subsection{Different stochastic interpretations}
\label{subsec:interps}
% Discuss Stratonovich and general alpha interpretations. The only real changes are that the f and g functions become different. 
Thus far, we have only discussed Ito-interpreted variables. In general, one can consider SDEs in the so-called $\alpha$-interpretation \cite{shi2012, yuan2012, tang_summing_2014}; the choice $\alpha = 0$ corresponds to the Ito interpretation, while the choice $\alpha = 1/2$ corresponds to the also popular Stratonovich interpretation. If our SDE (Eq. \ref{eq:1Dsde}) is $\alpha$-interpreted, the corresponding Fokker-Planck equation reads
\begin{equation} \label{eq:1DalphaFP}
\begin{split}
\frac{\partial P(x,t)}{\partial t} =& - \frac{\partial}{\partial x} \left[ \left( f(x) + \alpha g(x) g'(x) \right) P(x,t)  \right]  \\
&+ \frac{1}{2} \frac{\partial^2}{\partial x^2} \left[ g(x)^2 P(x,t)  \right] \ .
\end{split}
\end{equation}
If we like, we can rewrite this as 
\begin{equation}
\frac{\partial P(x,t)}{\partial t} = - \frac{\partial}{\partial x} \left[ \bar{f}(x) P(x,t)  \right] + \frac{1}{2} \frac{\partial^2}{\partial x^2} \left[ g(x)^2 P(x,t)  \right] 
\end{equation}
where $\bar{f}$ is the effective drift function. Since the form of the Fokker-Planck equation is exactly the same as before, the derivation can be carried out just as before---the only difference will be the replacement of $f$ with $\bar{f}$ in the final result. 

%MAY WONDER WHY THIS RESULT SEEMS SUPERFICIALLY DIFFERENT FROM OTHERS IN THE LITERATURE [CAN CITE SOME MATH PAPERS THAT GIVE FULL RESULT FROM WIKIPEDIA]. IN PARTICULAR NOTICE THE ABSENCE OF AN $f'$ TERM. THE ANSWER [CITE WEBER PG. 37, CUGLIANDOLO- 'RULES OF CALCULUS IN THE PATH INTEGRAL REPRESENTATION...'] HAS TO DO WITH THE CHOSEN DISCRETIZATION OF THE ACTION. OTHER DISCRETIZATIONS YIELD OTHER ACTIONS IN THE CONTINUOUS LIMIT. \\
%
%SURPRISING THING [DISCUSSED BY CUGLIANDOLO et al 'RULES...'] IS THAT NOT ONLY STOCHASTIC INTERPRETATION MATTERS, BUT CHOSEN WAY TO DISCRETIZE ACTION MATTERS TOO. CUGLIANDOLO ET AL SUMMARIZE: ``path integral extra sensitive to discretization issues'' \\

\subsection{SDEs with explicit time-dependence}

In the case of SDEs with explicit time-dependence, the derivation proceeds almost as it does in Sec. \ref{sec:1Dderivation}, but the notation becomes more complicated. When the functions $f$ and $g$ from our SDE (see Eq. \ref{eq:1Dsde}) have explicit time-dependence, the Hamiltonian (Eq. \ref{eq:Hopdef}) becomes explicitly time-dependent, which means its formal solution is
\begin{equation} \label{eq:formalslnTD}
\ket{\psi(t_f)} = \exp\left\{ \int_{t_0}^{t_f} \hat{H}(t) \ dt \right\} \ket{\psi(t_0)} 
\end{equation}
instead of Eq. \ref{eq:EOMformalsolution}. The series expansion corresponding to Eq. \ref{eq:formalslnTD} is often called the Magnus expansion \cite{magnus1954, blanes2009, dinh2017}, and is analogous to Dyson's series, which may be familiar from standard textbook treatments of quantum mechanics and quantum field theory \cite{dyson2006, schwartz2014, sakurai2017}.

The infinitesimal form of the propagator is in this case
\begin{equation}
\exp\left\{ \int_{t_j}^{t_j + \Delta t} \hat{H}(t) \ dt \right\} \approx 1 + \hat{H}(t_j) \Delta t \ ,
\end{equation}
and the rest of the derivation is otherwise the same. 

\subsection{Multiple variables}
\label{subsec:multivar}

% Write down form for multiple Ito variables, since the different interpretations only change the f and g functions.
To illustrate how one could derive a path integral as above for systems with multiple variables, consider a system of Ito-interpreted concentration-type SDEs (given the previous subsection, assuming some of the variables have a stochastic interpretation other than Ito is a trivial change). We have
\begin{equation} \label{eq:arbDsde}
d\mathbf{X} = \mathbf{f}(\mathbf{X}, t) dt + \sigma(\mathbf{X}, t) d\mathbf{W}
\end{equation}
where $\mathbf{X} = (X_1, ..., X_N)$ is an $N$-dimensional stochastic process, $\mathbf{W}$ is an $M$-dimensional Wiener process, $\mathbf{f}$ is $N$-dimensional, and $\sigma$ is $N \times M$. The corresponding Fokker-Planck equation reads
\begin{equation}
\begin{split}
\frac{\partial P(\mathbf{x},t)}{\partial t} =& - \sum_{i = 1}^N \frac{\partial}{\partial x^i} \left[ f^i P(\mathbf{x},t) \right] \\
&+ \sum_{i = 1}^N \sum_{j = 1}^N \frac{\partial^2}{\partial x^i \partial x^j} \left[ D^{ij} P(\mathbf{x},t) \right]
\end{split}
\end{equation}
where 
\begin{equation} \label{eq:difftensor}
D^{ij} := \frac{1}{2} \sum_{k = 1}^M \sigma^{ik} \sigma^{kj}
\end{equation}
is called the diffusion matrix. In Einstein's shorthand notation \cite{einstein1916}, where summation over repeated indices is implied, we can write
\begin{equation}
\frac{\partial P}{\partial t} = - \partial_i f^i P + \partial_i \partial_j D^{ij} P \ .
\end{equation}
The momentum operator and eigenkets (for $N$ concentration-type variables) generalize in the obvious way to
\begin{equation}
\begin{split}
\hat{p}_i &:= - \partial_i \\
\ket{p} &:= \int_0^{\infty} dx_1 \cdots \int_0^{\infty} dx_N \ e^{- \textbf{p} \cdot \textbf{x}} \ket{x} \\
1 &= \frac{1}{(2 \pi i)^N} \int_{-\infty}^{\infty} dp_1 \cdots \int_{-\infty}^{\infty} dp_N \ \ket{p} \bra{p} \ .
\end{split}
\end{equation}
In this case, the MSRJD-type (c.f. Eq. \ref{eq:MSRJDLagrangian}) result has
\begin{equation} \label{eq:multivarLagrangian}
L = i p_i (\dot{x}^i - f^i) + p_i D^{ij} p_j 
\end{equation}
and the corresponding Onsager-Machlup (c.f. Eq. \ref{eq:OMLagrangian}) result has
\begin{equation}
L = \frac{1}{4} (\dot{x}^i - f^i) D^{-1}_{ij} (\dot{x}^j - f^j) \ .
\end{equation}
Although we only discussed multiple Ito-interpreted concentration-type variables in this section, similar results can be derived for arbitrary numbers of variables with arbitrary combinations of variable domains.

\subsection{Explicitly incorporating a dimensionful control parameter}

In quantum mechanics, the dimensionful parameter $\hbar$ in some sense controls the validity of semiclassical approximations. In the path integral formulation, it does so through its appearance next to the action $S$ (c.f. Eq. \ref{eq:QMconfig}). Given that $\hbar$ is typically `small' relative to $S$, the exponential oscillates wildly, and paths for which the action $S$ is close to stationary (the `classical' paths) dominate \cite{feynman2010}. 

In some sense, the appearance of $\hbar$ in the argument of the path integral's exponential is due to its appearance in the canonical commutation relation
\begin{equation} \label{eq:QMccr}
[\hat{x}, \hat{p}]_{QM} = i \hbar \ ,
\end{equation}
or equivalently defining the momentum operator (via its action on states in the coordinate basis) to be
\begin{equation} \label{eq:QMmomentum}
\hat{p}_{QM} := - i \hbar \frac{\partial}{\partial x}  \  .
\end{equation}

Consider an Ito-interpreted concentration-type variable, as was considered in Sec. \ref{sec:1Dderivation}. Looking back through the derivation, we notice that we chose 
\begin{equation} \label{eq:Itomomentum}
\hat{p} := - \frac{\partial}{\partial x}  
\end{equation}
and
\begin{equation} \label{eq:Itoccr}
[\hat{x}, \hat{p}] = 1 \ ,
\end{equation}
i.e. the units of $p$ are chosen to be the inverse of the units of $x$, causing the RHS of Eq. \ref{eq:QMccr} to be dimensionless. If there is some important and dimensionful control parameter in one's problem, Eq. \ref{eq:Itomomentum} and Eq. \ref{eq:QMccr} can be amended to include it.

For example, in chemical kinetics, one can invoke the largeness of the system volume $\Omega$ as a way to approximate the chemical Kramers-Moyal equation as a more tractable Fokker-Planck equation \cite{kubo1973, gillespie2000, gardiner2009, vankampen2007}. In this scheme, the validity of the commonly used prescription for the deterministic rate equations can be understood as a kind of semiclassical approximation; just as the smallness of $\hbar$ in quantum mechanics is used to argue that classical behavior dominates, the largeness of $\Omega$ is used to argue that the deterministic rate equations well describe the system's dynamics.

Let us explicitly implement the system volume $\Omega$ as a dimensionful control parameter. Suppose we have a system described by the chemical Kramers-Moyal equation, and that we go from a dimensionless number variable $n$ to a dimensionful concentration variable $x := n/\Omega$. Now suppose that we have truncated the Kramers-Moyal equation at second order to obtain a Fokker-Planck equation. Gillespie showed \cite{gillespie1976, gillespie1992, gillespiebook1992, gillespie2000} that the volume-dependence of the propensity functions for monomolecular, bimolecular, and trimolecular reactions comes out in just such a way that the Fokker-Planck equation for the modified system can be written
\begin{equation} \label{eq:VFP}
\frac{\partial P(x,t)}{\partial t} = - \frac{\partial}{\partial x} \left[ f(x) P(x,t)  \right] + \frac{1}{2} \frac{1}{\Omega} \frac{\partial^2}{\partial x^2} \left[ g(x)^2 P(x,t)  \right] \ .
\end{equation}
We can define the momentum operator as
\begin{equation} \label{eq:Vmomentum}
\hat{p} := - \frac{1}{\Omega} \frac{\partial}{\partial x}   \ ,
\end{equation}
which yields the canonical commutation relation
\begin{equation} \label{eq:Vccr}
[\hat{x}, \hat{p}] = \frac{1}{\Omega} \ .
\end{equation}
The momentum eigenkets are\footnote{As in Sec. \ref{subsubsec:peigenket}, these are only uniquely defined up to a multiplicative constant. Changing the constant in the definition of these eigenkets just changes the constant that appears in front of the resolution of the identity.}
\begin{equation} \label{eq:Vpket}
\ket{p} := \int_0^{\infty} dx \ e^{- p \Omega x} \ket{x} \ .
\end{equation}
and with them one can construct the operator
\begin{equation} \label{eq:VROIp}
1 = \frac{\Omega}{2\pi i} \int_{-i \infty}^{i \infty} dp \ \ket{p} \bra{p} \ .
\end{equation}

The Hilbert space formulation of this problem (Eq. \ref{eq:VFP}) then becomes 
\begin{equation} \label{eq:VgfEOM}
\frac{\partial \ket{\psi(t)}}{\partial t}  = \Omega \hat{H} \ket{\psi} \ ,
\end{equation}
where the Hamiltonian $\hat{H}$ is defined the same way as before (Eq. \ref{eq:Hopdef}). We again have the formal solution
\begin{equation} \label{eq:VEOMformalsolution}
\ket{\psi(t_f)} = e^{\Omega \hat{H} (t_f - t_0)} \ket{\psi(t_0)} \ .
\end{equation}
From here, everything is the same except for the final result (c.f. Eq. \ref{eq:discreteMSRJD}), which reads
\begin{widetext} 
\begin{equation} \label{eq:VdiscreteMSRJD}
\begin{split}
P &= \lim_{N \to \infty} \left( \frac{\Omega}{2 \pi} \right)^N \int dp_N \prod_{j = 1}^{N-1}  dx_j dp_j   \exp{- \Omega \sum_{j = 1}^N \left[ i p_j \left( \frac{x_j - x_{j-1}}{\Delta t} - f(x_{j-1}) \right) + \frac{1}{2} {p_j}^2 g(x_{j-1})^2 \right] \Delta t}  \\
&= \int \mathcal{D}[x(t)] \mathcal{D}[p(t)] \ \exp\left\{- \Omega \ S[x, p] \right\} \ .
\end{split}
\end{equation}
\end{widetext}
In this case, the action $S$ is the same as before (c.f. Eq.  \ref{eq:discreteMSRJD}), but with extra factors of $\Omega$. The corresponding Onsager-Machlup result is the same except for the replacement $g^2 \to g^2/\Omega$.

There are two ways to account for the `thermodynamic limit' of Eq. \ref{eq:VdiscreteMSRJD}, in which the number of molecules $n$ and the system volume $\Omega$ are both taken to infinity such that their ratio $x$ remains constant \cite{gillespie2000}: (i) take $\Omega \to \infty$, and use Laplace's method \cite{laplace1986, erdelyi1956, wojdylo2006} to argue that the least action path dominates; or (ii) change variables to $q_j = \Omega p_j$ for all $j$, and note that the noise term in the action vanishes as we take $\Omega \to \infty$. 

%take $\Omega$ to infinity in the original formalism, where $g^2 \to g^2/\Omega$, so that the noise term goes away in the path integral action. \\

As an aside, it is amusing to note that considering diffusion-type variables (see Sec. \ref{ssec:diffandcompact}) with a dimensionful control parameter like $\Omega$ yields a formalism (momentum operator, canonical commutation relation, momentum eigenket, momentum resolution of the identity) that looks very much like quantum mechanics, but with $1/\Omega$ playing the role of $\hbar$:
\begin{equation}
\label{eq:diffplusV}
\begin{split}
\hat{p} &:= - \frac{i}{\Omega} \frac{\partial}{\partial x} \\
[\hat{x}, \hat{p}] &= \frac{i}{\Omega} \\
\ket{p} &:= \int_{-\infty}^{\infty} dx \ e^{i \Omega p x} \ket{x} \\
1 &= \frac{\Omega}{2\pi} \int_{-\infty}^{\infty} dp \ \ket{p} \bra{p} \ .
\end{split}
\end{equation}

In summary, our formalism offers a natural way to understand large volume approximations, their thermodynamic limits, and their relationship with ordinary quantum mechanics.

\subsection{Other PDEs first order in time}

One nice feature of this derivation is that it does not explicitly invoke the Markov property of these processes; indeed, one can imagine `turning the crank' and using this method to find formal solutions to any linear partial differential equation which is first order in time\footnote{An approach with this feature is not new; Graham used a similar idea \cite{graham1977} to derive the quantum mechanical and Fokker-Planck path integrals in the same paper.}.

For example, one relevant PDE in the study of gene regulation and chemical kinetics is the chemical Kramers-Moyal equation \cite{gillespie2000, kramers1940, moyal1949}, which is an approximation to the CME. For a system with $N$ species and $M$ reactions, it reads
\begin{widetext}
\begin{equation}
\label{eq:km}
\frac{\partial P(\mathbf{x},t)}{\partial t} = \sum_{n = 1}^{\infty} (-1)^n \sum_{m_1 + \cdots + m_N = n} \frac{1}{m_1! \cdots m_N!}  \frac{\partial^n}{\partial x_1^{m_1} \cdots \partial x_N^{m_N}} \left\{ \left[ \sum_{j = 1}^M (\nu^{j1})^{m_1} \cdots (\nu^{jN})^{m_N} a_j(\mathbf{x}) \right]  P(\mathbf{x},t)  \right\} \ .
\end{equation}
\end{widetext}
%\begin{equation} \label{eq:km}
%\begin{split}
%& \frac{\partial P(\mathbf{x},t)}{\partial t} \\
%=& \sum_{n = 1}^{\infty} (-1)^n \sum_{m_1 + \cdots + m_N = n} \frac{1}{m_1! \cdots m_N!} \\
%&\times \frac{\partial^n}{\partial x_1^{m_1} \cdots \partial x_N^{m_N}} \left\{ \left[ \sum_{j = 1}^M (\nu^{j1})^{m_1} \cdots (\nu^{jN})^{m_N} a_j(\mathbf{x}) \right]  P(\mathbf{x},t)  \right\} \ .
%\end{split}
%\end{equation}
where $\mathbf{x} := (x_1, ..., x_N)$ is the (nonnegative) concentration of each species, $\nu^{ij}$ is the reaction stoichiometry matrix, and the $a_j$ are the propensity functions. We will use our formalism to derive a formal path integral solution to this equation.

While the fact that this PDE has an infinite number of terms may first seem daunting, the calculation turns out to be no more complicated than the multivariable calculation from Sec. \ref{subsec:multivar}. Define
\begin{equation}
T_{m_1 \cdots m_N}  := \frac{1}{m_1! \cdots m_N!} \sum_{j = 1}^M (\nu^{j1})^{m_1} \cdots (\nu^{jN})^{m_N} a_j
\end{equation}
so we can write Eq. \ref{eq:km} as
\begin{equation}
\frac{\partial P}{\partial t} = \sum_{n = 1}^{\infty} (-1)^n \ \sum_{m_1 + \cdots + m_N = n} \partial_1^{m_1} \cdots \partial_N^{m_N} \ T_{m_1 \cdots m_N} P \ .
\end{equation}
Following the procedure outlined in the previous sections, we can find a formal path integral solution with Lagrangian
\begin{equation} \label{eq:KMLagrangian}
\begin{split}
L =& \ i \ p_i \dot{x}^i \\
&- \sum_{n=1}^{\infty} \ \sum_{m_1 + \cdots + m_N = n} (i p_1)^{m_1} \cdots (i p_N)^{m_N} \ T_{m_1 \cdots m_N} \ .
\end{split}
\end{equation}
While the infinite number of terms in the action suggests that the mathematical content of the above expression is dubious, it may be useful if it is handled with care. As a sanity check, observe that truncating this Lagrangian at $n = 2$ yields
\begin{equation} \label{eq:KMLagrangiantruncated}
L = i \ p_i (\dot{x}^i - f^i) + p_i D^{ij} p_j \ ,
\end{equation}
with
\begin{equation} \label{eq:KMtruncationdata}
\begin{split}
f^i &= \sum_{k = 1}^M \nu^{ki} a_k \\
D^{ij} &= \frac{1}{2} \sum_{k = 1}^M \nu^{ki} \nu^{kj} \ a_k  \ ,
\end{split}
\end{equation}
i.e. exactly the result for the corresponding multivariable chemical Langevin equation (c.f. Eq. \ref{eq:multivarLagrangian} and see Eq. 24 of Gillespie \cite{gillespie2000}). 

There is no Onsager-Machlup-type result corresponding to Eq. \ref{eq:KMLagrangian}, because it is not generally quadratic in the $p_i$ variables, so they cannot easily be integrated out.

%=======================================================================

\section{Discussion}
\label{sec:discussion}

We have developed a quantum mechanics-like formalism for deriving a path integral description of systems described by SDEs (or equivalently, by a Fokker-Planck equation). Our approach accounts for different variable domains, avoids the unwieldy Jacobian calculations involved with the infinitesimal transition probability approach, does not encounter discretization issues when treating systems with different stochastic interpretations, easily generalizes beyond the Fokker-Planck equation to other PDEs of interest that are first order in time, works for a wide variety of systems, and makes the relationship between Fokker-Planck stochastic dynamics and quantum mechanics clearer.

To our knowledge, we are the first to point out the relevance of variable domain in constructing a stochastic path integral formalism (and in particular, in defining the appropriate momentum eigenkets). The compact-type variable path integral we introduced here may better treat confined diffusion problems---for example, diffusion near a cellular border \cite{holmes2019}---while a diffusion-type variable path integral may better treat bulk diffusion problems. The idea that moving to the momentum space probability density corresponds to taking a domain-appropriate transform (Laplace, Fourier, or Fourier series) can in principle be generalized to other transforms we did not discuss, which presumably would be associated with their own path integral formalisms. 

Although our main results (Eq. \ref{eq:discreteMSRJD} and Eq. \ref{eq:discreteOM}) may appear superficially different from some others used in the gene regulation literature \cite{wang2011, tang_summing_2014, zhouli2016, li2014, li2016landscape, li2013, wang2010, hertz2018} (note the extra $f'$ term in many path integrals), we stress that difference is due to choosing different \textit{discretizations} for the action\footnote{See pg. 36-37 of Weber and Frey \cite{weber_master_2017} for an elementary discussion, and Cugliandolo et al. \cite{cugliandolo2017, cugliandolo2019} for more detailed discussions of this phenomenon.}. We choose the discretization where functions $h(x)$ are approximated as $h(x_j)$, while some others choose to approximate them as $h(\bar{x_j})$, where $\bar{x_j} := \kappa x_{j+1} + (1-\kappa) x_j$ for some $0 \leq \kappa \leq 1$. Somewhat confusingly, this choice of discretization is independent from the choice of stochastic interpretation for the underlying Langevin equation/SDE; one may choose a Stratonovich interpretation, for example, but evaluate functions within the discretized action at $x_j$ instead of at $\frac{1}{2} x_{j+1} + \frac{1}{2} x_j$. We do such a thing here: we evaluate at $x_j$ regardless of the chosen interpretation, although one can recover the other path integrals by starting with ours and doing an appropriate change of variables. 

%There is a minor controversy regarding the proper form of the Lagrangian (Eq. \ref{eq:OMLagrangian} is our result); Wang et al. use one result in the gene regulation literature, Ao et al. use another result. Our method does not assume any particular stochastic interpretation (see Sec. \ref{subsec:interps})...and avoids the many pitfalls associated with changes of variables, as Tang et al. use in their derivation. \\
%
%Our method agrees with the result of Cugliandolo et al. \\

As suggested in the introduction, our approach clarifies the link between the Doi-Peliti path integral \cite{peliti1985} and the Fokker-Planck path integral. In particular, it is likely one can derive the Fokker-Planck description from the Doi-Peliti description either (i) at the path integral level, by approximating the action; or (ii) at the \textit{formalism} level, by approximating the Hamiltonian and rewriting the relevant operators.

Because our derivation does not assume anything about noise (i.e. the function $g$ in Eq. \ref{eq:1Dsde}, or equivalently the diffusion tensor $D^{ij}$ for multivariable systems like Eq. \ref{eq:arbDsde}), we can construct path integral descriptions of systems with strong and state-dependent/multiplicative noise. Although Gillespie showed \cite{gillespie2000} some time ago that chemical Langevin equations generically have multiplicative noise terms, the qualitative impact of these terms (and of intrinsic noise more generally) on gene regulation models \cite{holmes2017,wang2017cell, ghosh2012, sharma2017, zhang2017, newby2012, newby2015} is only recently being appreciated.

\section{Conclusion}
\label{sec:conclusion}

We have presented a new formalism for deriving stochastic path integrals associated with the Fokker-Planck equation, and this formalism more closely parallels the derivations usually used in quantum mechanics and Doi-Peliti field theory. It is fairly general, given that it can treat variables with different domains, different stochastic interpretations, explicit time-dependence, arbitrary numbers of variables, dimensionful control parameters, and any PDE first order in time. We hope that this formalism helps make the mathematical analogies  between stochastic mechanics and quantum mechanics more transparent.

\begin{acknowledgments}
This work was supported by NSF Grant \# DMS 1562078.
\end{acknowledgments}

\bibliography{pathint_bib, pathint_bib2, apps}

% Create the reference section using BibTeX:
%\bibliography{basename of .bib file}

\end{document}